# The rotating wind of the quasar PG 1700+518

S. Young[1,2], D. J. Axon[1,2], A. Robinson[1,2], J. H. Hough[2] & J. E. Smith[2]

[1]Department of Physics, Rochester Institute of Technology, 84 Lomb Memorial Drive, Rochester, New York 14623, USA. [2]Centre for Astrophysics Research, Science & Technology Research Institute, University of Hertfordshire, Hatfield AL10 9AB, UK.

**It is now widely accepted that most galaxies undergo an active phase, during which a central super-massive black hole generates vast radiant luminosities through the gravitational accretion of gas[1,2]. Winds launched from a rotating accretion disk surrounding the black hole are thought to play a critical role, allowing the disk to shed angular momentum that would otherwise inhibit accretion[3,4]. Such winds are capable of depositing large amounts of mechanical energy in the host galaxy and its environs, profoundly affecting its formation and evolution[5–7], and perhaps regulating the formation of large-scale cosmological structures in the early Universe[8,9]. Although there are good theoretical grounds for believing that outflows from active galactic nuclei originate as disk winds[10], observational verification has proven elusive. Here we show that structures observed in polarized light across the broad H$\alpha$ emission line in the quasar PG 1700+158 originate close to the accretion disk in an electron scattering wind. The wind has large rotational motions (~4,000 km s$^{-1}$), providing direct observational evidence that outflows from active galactic nuclei are launched from the disks. Moreover, the wind rises nearly vertically from the disk, favouring launch mechanisms that impart an initial acceleration perpendicular to the disk plane.**

The blueshifted broad absorption lines (BALs) observed in the spectra of quasars provide some of the most compelling evidence for powerful outflows from active galactic nuclei (AGN)[11]. Although only about 15% of quasars exhibit BALs[12], it is generally believed that the high velocity winds in which these features are formed are ubiquitous[13], the BALs being observed only when our line of sight happens to intercept the wind, which occupies a relatively small solid angle.

Our optical spectropolarimetry of the low-redshift ($z = 0.29$) BAL quasar PG 1700+518 (ref. 14) shows significant variations in both the degree ($p(\lambda)$) and position angle ($\theta(\lambda)$) of polarization across both the broad Balmer H$\alpha$ and H$\beta$ emission lines (Fig. 1). In polarized flux ($p(\lambda) \times F(\lambda)$), the broad Balmer line profiles are significantly redshifted relative to their total flux counterparts. Relative to its value in the continuum, $\theta(\lambda)$ deviates by 30–40°, rotating in opposite directions in the blue and red wings of the line profile, with the reversal occurring near the central wavelength of the line. These polarization structures can be discerned in earlier observations[15,16], but have, hitherto, been left unexplained.

It is generally accepted that the broad Balmer lines in AGN are predominantly emitted by the accretion disk itself[17]. Polarization position angle rotations similar to those seen in PG 1700+518 are frequently observed in Seyfert 1 galaxies, in some of which the dominant polarization mechanism is believed to be scattering of emission line photons from the accretion disk in an equatorial scattering medium co-planar with, and closely surrounding, the disk. The characteristic $\theta(\lambda)$ variation across the line profile naturally arises from the rotational motion of the disk, as long as the angular size of the disk as seen by the scattering medium is sufficiently large[18]. In the Seyfert galaxies, however, the scattered line profiles do not have the large redshift seen in PG 1700+518. This net Doppler shift implies that the gas emitting the lines and the scattering electrons are moving apart with a velocity of ~4,000 km s$^{-1}$.

In contrast to the Seyfert galaxies, therefore, the optical polarization spectrum of PG 1700+518 exhibits strong signatures of both rotational and radial motions. The most plausible interpretation of the redshifted scattered line profiles is that the scattering medium is part of a high velocity outflow. On the other hand, the rotation in $\theta(\lambda)$ could be produced by rotational motions in the disk, or in the scattering region, or, as we argue below, both. Furthermore, the large amplitude of the $\theta(\lambda)$ swing implies that the scattering occurs close to the line-emitting region of the disk. The polarization spectrum, therefore, is consistent with scattering in a wind expanding away from the accretion disk. The observations cannot, however, be explained simply by adding a radial expansion component to the equatorial scattering region invoked to explain the broad Hα polarization in Seyfert galaxies (effectively making it an equatorial wind).

We have used an extensive grid of three-dimensional scattering calculations, described in detail elsewhere[19], to explore emission-line polarization signatures resulting from scattering in winds having different geometries, launched from a range of radii (relative to the Balmer-line emitting zone) within the disk. On quite general grounds, we infer that the scattering particles are electrons in ionized gas whose temperature is less than a few times $10^5$ K. Scattering by dust grains can be ruled out, as the large $\theta(\lambda)$ rotation requires that the scattering takes place within a radius of $\sim 10^{16}$ m (0.3 pc) of the central AGN, well inside the radius ($\sim$1–2 pc for this object) within which unshielded grains will be destroyed by the quasar's radiation field[20]. On the other hand, if the temperature of the scattering medium is $\gg 10^5$ K, thermal Doppler broadening would smear out the polarization structure (Supplementary Fig. 1).

The model grid allows us to determine the essential elements such a scattering wind must possess in order to produce the principal features of the Hα polarization, namely, a large amplitude rotation in $\theta(\lambda)$, anchored near the line centre, and the observed redshift in polarized flux. These features can be produced simultaneously only by models (Fig. 2, see also Supplementary Fig. 2) in which the wind rises vertically from the disk in a cylindrical configuration and includes both rotational ($v_{rot}$) and vertical ($v_z$) velocity components. Conical outflows naturally produce a redshifted line profile in polarized flux, but also a redshifted $\theta(\lambda)$ rotation. For equatorial winds, on the other hand, the $\theta(\lambda)$ rotation is both redshifted and of low amplitude. Furthermore, in order for the scattered line profile to have a Doppler width and redshift comparable to those observed, both the rotational and vertical velocity components of the cylindrical wind must have magnitudes comparable to the local keplerian velocity of the disk, $v_K$ (Supplementary Fig. 3). The vertical extent ($h$) of the scattering region above the disk is also strongly constrained by the observed polarization spectra. In order to produce a redshifted scattered line profile, while maintaining a large amplitude rotation in $\theta(\lambda)$, we require $2r_{in} < h \leq 4r_{in}$, where $r_{in}$ is the inner radius of the scattering region in cylindrical coordinates (Supplementary Fig. 4). For similar reasons, the wind must be launched from approximately the same annular region within the disk as the Balmer-line emitting gas (Supplementary Fig. 5). Finally, values of $p(\lambda)$ and the $\theta(\lambda)$ rotation amplitude consistent with the observations are produced when the system is viewed at an intermediate inclination ($i \approx 45°$).

The radius of the Balmer-line emitting zone in PG 1700+518 has been determined by reverberation mapping to be $\sim 7 \times 10^{15}$ m. Combining this radius with the Doppler velocity dispersion inferred from the width of the line profile leads (under the assumption that the emitting gas is gravitationally bound) to a mass for the central black hole of $\sim 8 \times 10^8\ M_\odot$ (ref. 21; here $M_\odot$ is the solar mass). For this mass, the keplerian velocity at the launch radius of the wind is $v_K \approx 4{,}000$ km s$^{-1}$. Our models imply that $v_z \approx v_{rot} \approx v_K \approx 4{,}000$ km s$^{-1}$. This velocity is small compared to the outflow speeds characteristic of the BALs (1–2 $\times 10^4$ km s$^{-1}$ in PG 1700+518; ref. 14), which implies that the scattering takes place in a region where the wind has not yet reached its terminal velocity, or where the acceleration mechanism is less efficient than in the BAL-producing zone.

Theoretical studies have shown that disk winds driven either by radiation-pressure or magneto-centrifugal forces, or some combination thereof, are capable of producing velocities and mass-flow rates consistent with the observed properties of BALs[22–24]. Our interpretation of the spectropolarimetry data is consistent with a scenario in which the wind is launched from a relatively narrow annulus within the disk, which includes the Balmer-line emitting region, and is initially accelerated vertically (Fig. 3). The velocity field at the base of the wind, with azimuthal and vertical components dominating the radial component, favours models in which the initial acceleration is due to thermal[25] or magnetic pressure[26] or radiation pressure of photons emitted locally by the disk[10,27]. On the other hand, our results are problematic for models invoking equatorial winds[4,22], which require initial streamlines at large angles to the disk normal (for example, >30° for a wind launched by magneto-centrifugal forces).

A viable radiatively driven disk wind must be shielded from the high energy continuum source of the AGN to prevent the gas becoming over-ionized, as this would eliminate line absorption, the major source of the radiation pressure driving force[22]. Theoretical work on this problem suggests that disk winds are self-shielded by an inner zone, which is too highly ionized for efficient line driving but which is effectively opaque to X-rays[24,27]. The scattering region can be plausibly identified with such an X-ray-shielding inner wind zone. Our polarimetry data do not directly constrain the region in which the BALs arise. However, in such two-zone wind models, these features are formed in the shielded outer wind[27]. Here radiative acceleration due to ultraviolet line opacity is efficient, and the streamlines, although nearly vertical close to the disk, become effectively radial at higher elevations (Fig. 3).

It has been argued that disk winds explain many observed properties of AGN, including BALs, the X-ray 'warm absorber' and ultraviolet absorption lines[28]. The spectropolarimetric properties of PG 1700+518 provide compelling evidence not only that such disk winds exist but also that they are a key element in any unified picture of the inner structure of AGN. More specifically, our results demonstrate that the polarization structure across the broad emission lines constrains the geometry, launch site and kinematics of the scattering medium, and therefore places vital and unique observational constraints on physical models of disk winds. Furthermore, these insights into the nature of AGN disk winds may also be applicable to a wide variety of other astrophysical sources in which keplerian disks are thought to generate outflows, such as young stellar objects, or B[e] stars. One of the latter, in particular, exhibits emission line polarization structures reminiscent of those observed in PG 1700+518 (ref. 29).

1. Rees, M. J. Black hole models for active galactic nuclei. *Annu. Rev. Astron. Astrophys.* **22,** 471–506 (1984).
2. Ferrarese, L. & Ford, H. Supermassive black holes in galactic nuclei: Past, present and future research. *Space Sci. Rev.* **116,** 523–624 (2005).
3. Crenshaw, D. M., Kraemer, S. B. & George, I. M. Mass loss from the nuclei of active galaxies. *Annu. Rev. Astron. Astrophys* **41,** 117–167 (2003).
4. Blandford, R. D. & Payne, D. G. Hydromagnetic flows from accretion discs and the production of radio jets. *Mon. Not. R. Astron. Soc.* **199,** 883–903 (1982).

5. Silk, J. & Rees, M. J. Quasars and galaxy formation. *Astron. Astrophys.* **331,** L1–L4 (1998).

6. Scannapieco, E., Silk, J. & Bouwens, R. AGN feedback causes downsizing. *Astrophys. J.* **635,** L13–L16 (2005).

7. Di Matteo, T., Springel, V. & Hernquist, L. Energy input from quasars regulates the growth and activity of black holes and their host galaxies. *Nature* **433,** 604–607 (2005).

8. Bower, R. G. *et al.* Breaking the hierarchy of galaxy formation. *Mon. Not. R. Astron. Soc.* **370,** 645–655 (2006).

9. Menci, N., Fontana, A., Giallongo, E., Grazian, A. & Salimbeni, S. The abundance of distant and extremely red galaxies: The role of AGN feedback in hierarchical models. *Astrophys. J.* **647,** 753–762 (2006).

10. Shlosman, I., Vitello, P. A. & Shaviv, G. Active galactic nuclei — Internal dynamics and formation of emission clouds. *Astrophys. J.* **294,** 96–105 (1985).

11. Turnshek, D. A. Properties of the broad absorption-line QSOs. *Astrophys. J.* **280,** 51–65 (1984).

12. Reichard, T. A. *et al.* Continuum and emission-line properties of broad absorption line quasars. *Astron. J.* **126,** 2594–2607 (2003).

13. Weymann, R. J., Morris, S. L., Foltz, C. B. & Hewett, P. C. Comparisons of the emission-line and continuum properties of broad absorption line and normal quasi-stellar objects. *Astrophys. J.* **373,** 23–53 (1991).

14. Pettini, M. & Boksenberg, A. PG 1700+518 — A low-redshift, broad absorption line QSO. *Astrophys. J.* **294,** L73–L78 (1985).

15. Schmidt, G. D. & Hines, D. C. The polarization of broad absorption line QSOs. *Astrophys. J.* **512,** 125–135 (1999).

16. Ogle, P. M. *et al.* Polarization of broad absorption line QSOs. I. A spectropolarimetric atlas. *Astrophys. J. Suppl. Ser.* **125,** 1–34 (1999).

17. Rokaki, E. & Boisson, C. Consistency of accretion discs with Seyfert 1 UV fluxes and Hβ emission-line profiles. *Mon. Not. R. Astron. Soc.* **307,** 41–54 (1999).

18. Smith, J. E., Robinson, A., Young, S., Axon, D. J. & Corbett, E. A. Equatorial scattering and the structure of the broad-line region in Seyfert nuclei: Evidence for a rotating disc. *Mon. Not. R. Astron. Soc.* **359,** 846–864 (2005).

19. Young, S., Robinson, A., Axon, D. J. & Hough, J. H. Polarization signatures of disc-winds. *Astrophys. J.* (submitted).

20. Barvainis, R. Hot dust and the near-infrared bump in the continuum spectra of quasars and active galactic nuclei. *Astrophys. J.* **320,** 537–544 (1987).

21. Peterson, B. M. *et al.* Central masses and broad-line region sizes of active galactic nuclei. II. A homogeneous analysis of a large reverberation-mapping database. *Astrophys. J.* **613,** 682–699 (2004).

22. Murray, N., Chiang, J., Grossman, S. A. & Voit, G. M. Accretion disk winds from active galactic nuclei. *Astrophys. J.* **451,** 498–509 (1995).

23. de Kool, M. & Begelman, M. C. Radiation pressure-driven magnetic disk winds in broad absorption line quasi-stellar objects. *Astrophys. J.* **455,** 448–455 (1995).

24. Proga, D., Stone, J. M. & Kallman, T. R. Dynamics of line-driven disk winds in active galactic nuclei. *Astrophys. J.* **543,** 686–696 (2000).

25. Begelman, M. C., McKee, C. F. & Shields, G. A. Compton heated winds and coronae above accretion disks. I Dynamics. *Astrophys. J.* **271,** 70–88 (1983).

26. Stone, J. M. & Norman, M. L. Numerical simulations of magnetic accretion disks. *Astrophys. J.* **433,** 746–756 (1994).

27. Proga, D. & Kallman, T. R. Dynamics of line-driven disk winds in active galactic nuclei. II. Effects of disk radiation. *Astrophys. J.* **616,** 688–695 (2004).

28. Elvis, M. A Structure for quasars. *Astrophys. J.* **545,** 63–76 (2000).

29. Oudmaijer, R. D., Proga, D., Drew, J. E. & de Winter, D. The evolved B[e] star HD 87643: Observations and radiation driven disk-wind model for B[e] stars. *Mon. Not. R. Astron. Soc.* **300,** 170–182 (1998).

30. Young, S. A generic scattering model for AGN. *Mon. Not. R. Astron. Soc.* **312,** 567–578 (2000).

**Supplementary Information** is linked to the online version of the paper at www.nature.com/nature.

**Acknowledgements** This work is based on observations made with the William Herschel Telescope operated on the island of La Palma by the Isaac Newton Group in the Spanish Observatorio del Roque de los Muchachos of the Instituto de Astrofísica de Canarias. This research has made use of NASA's Astrophysics Data System. We acknowledge financial support from the Science and Technology Facilities Council, UK.

**Author Contributions** All authors contributed extensively to the work presented in this paper.

**Author Information** Reprints and permissions information is available at www.nature.com/reprints. Correspondence and requests for materials should be addressed to S.Y. (sxysps@rit.edu).

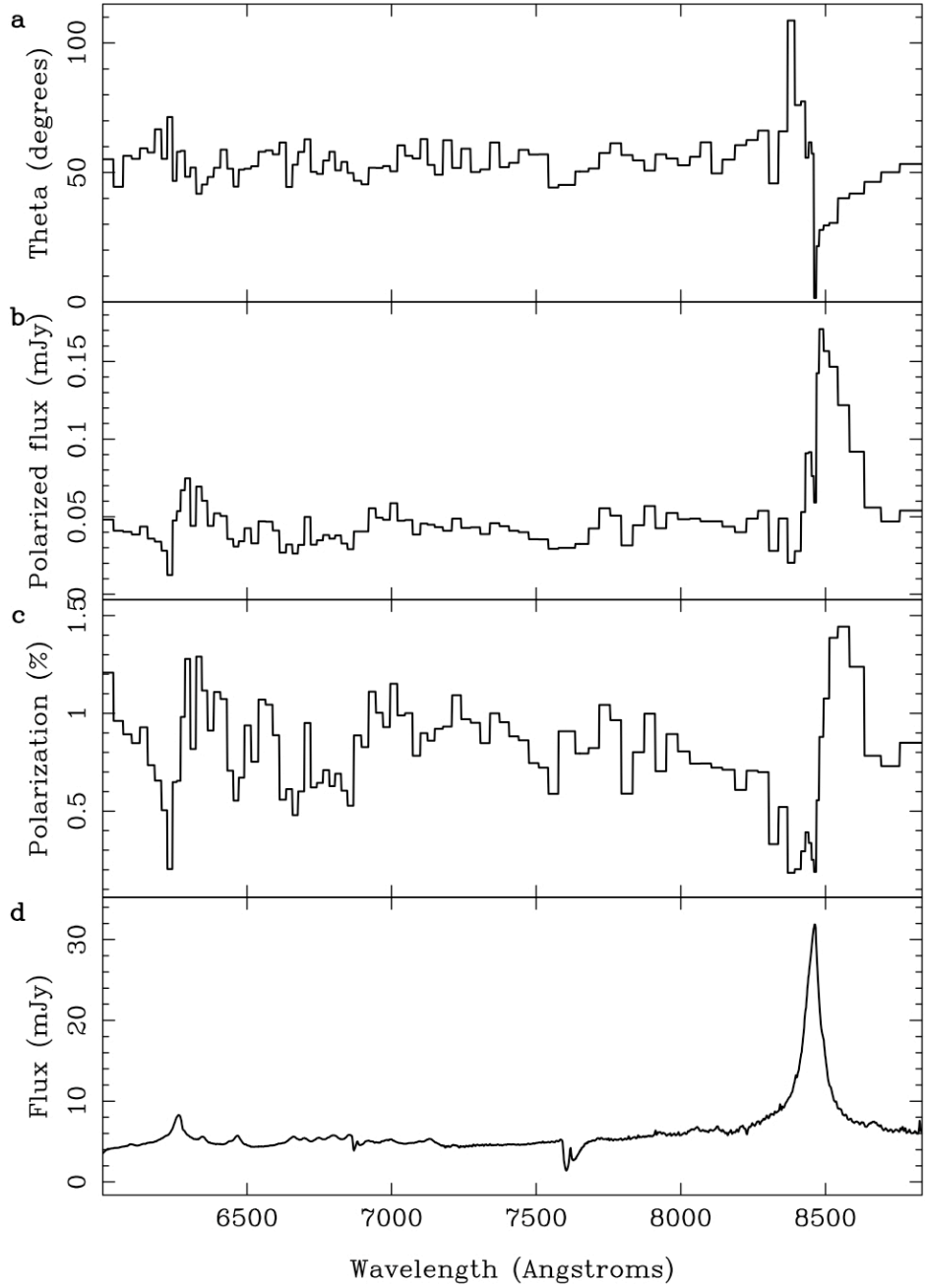

**Figure 1 The polarization data for the BAL quasar PG1700+518.** The data were obtained using the William Herschel Telescope in conjunction with the ISIS spectrograph,

at a spectral resolution of 3.4 Å, on the nights of 27 and 28 June 2003. The polarization data have been re-sampled into bins with an error of 0.1% in degree of polarization. **a**, The position angle of polarization, $\theta$; **b**, the polarized flux spectrum in mJy; **c**, the degree of polarization as a percentage; and **d**, the total flux spectrum in mJy. Note the large change in polarization position angle across the broad H$\alpha$ emission line, which reverses direction at the line peak. In polarized flux, the broad H$\alpha$ emission line is redshifted with respect to the wavelength of the peak in total flux.

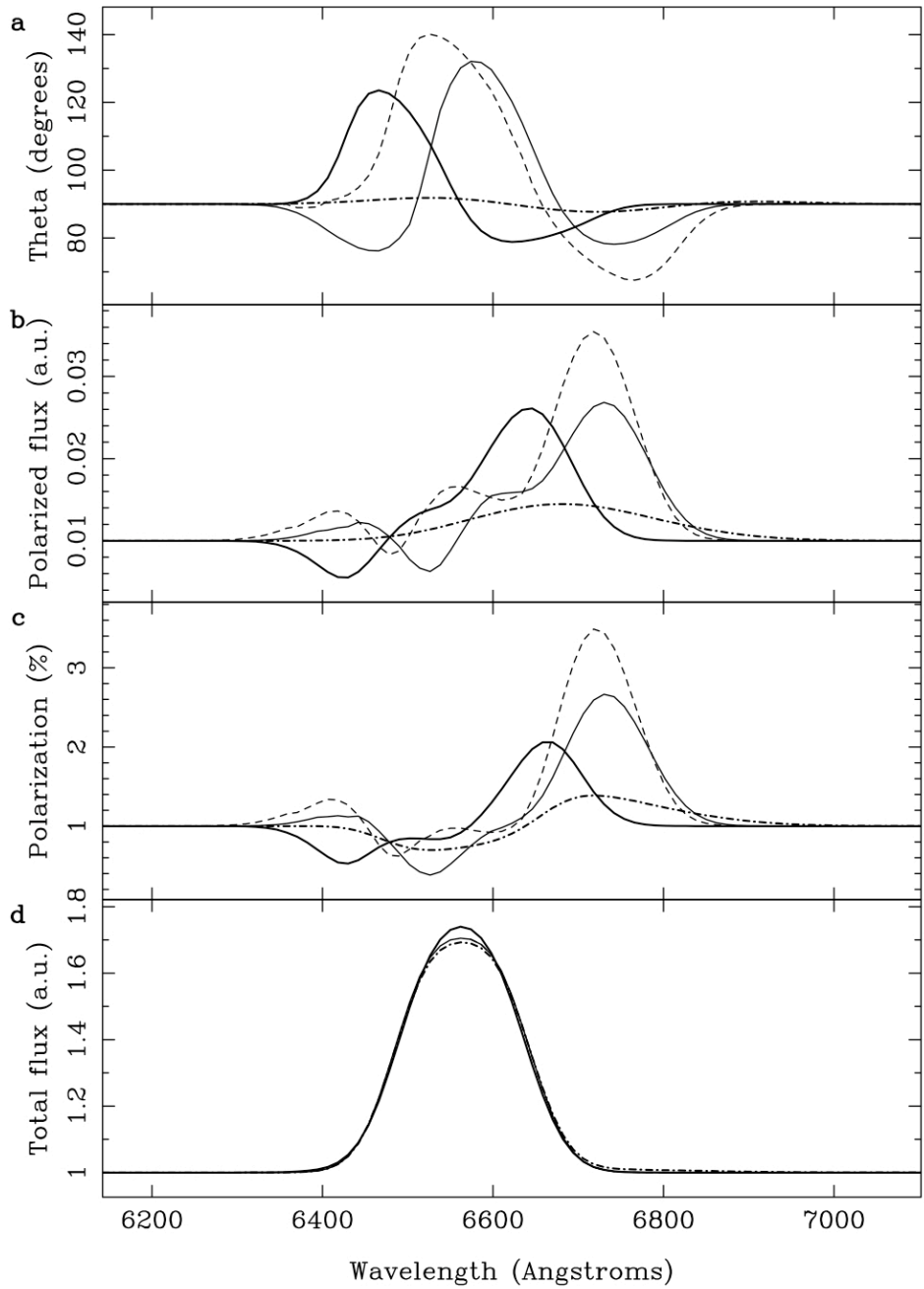

**Figure 2 Comparison of simulated polarization spectra for different scattering geometries.** The models were computed using our three-dimensional scattering code[30],

modified to include cylindrical and conical scattering winds. In each case, an emission line emitted by a quasar accretion disk is subsequently scattered in a wind. The wind is launched from the annular region of the accretion disk that produces the line emission. The models presented are: a cylindrical wind with both rotational and vertical outflow velocity components (solid heavy line); a disk-like, rotating, equatorial outflow (dot-dash heavy line); and a hollow-funnel-like conical wind with radial outflow only (solid lighter line) and with radial outflow combined with rotation (dashed lighter line). The inclination of the system polar axis to the line of sight is 48°, the outflow velocity is $0.02c$ and where applicable, the winds are rotating at the local keplerian velocity. The position angle of polarization ($\theta$; **a**) is measured relative to the projection of the system axis on the sky plane. The median $\theta$ for the equatorial outflow model is 0°, but has been shifted by 90° to ease comparison with the other models. The total flux spectrum (**d**) and the polarized flux spectrum (**b**) are in arbitrary units (a.u.); the degree of polarization (%) is shown in **c**. The equatorial outflow produces a red-shifted broad line in polarized flux but only a small variation in $\theta$ across the line. Both the cylindrical and conical wind models produce large-amplitude rotations of $\theta$, but only for the cylindrical wind is the rotation centred at the peak of the total flux line profile, as required. A colour version of this figure is included as Supplementary Information.

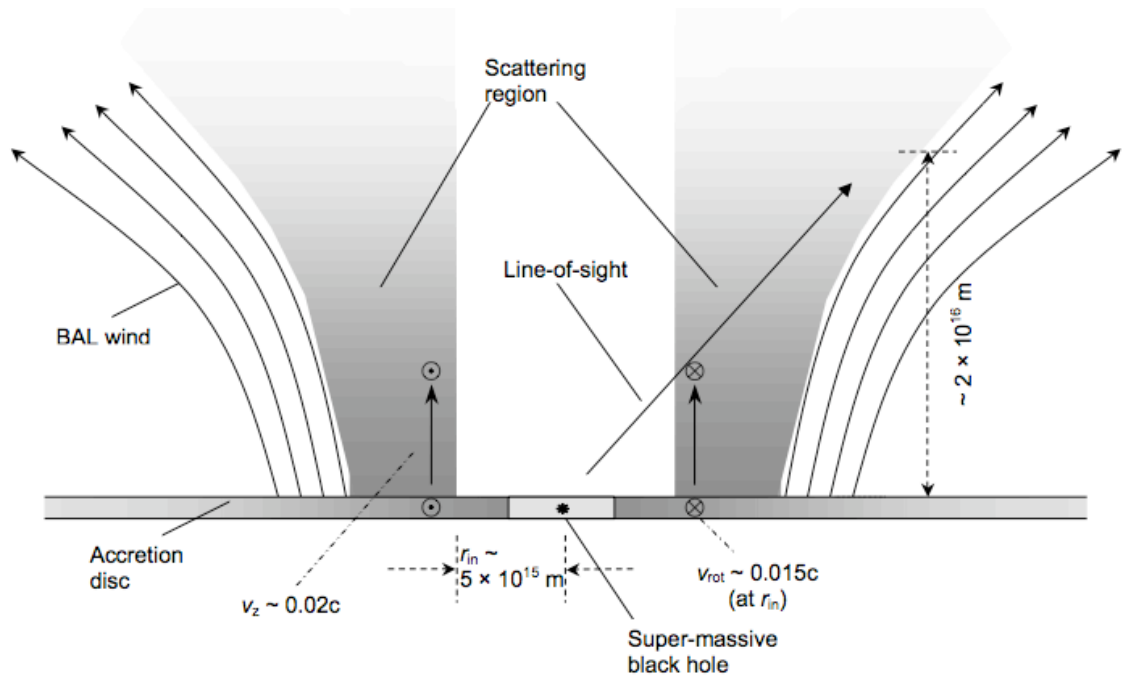

**Figure 3 A cross-section schematic illustration of the wind geometry inferred from the polarization and spectroscopic properties of PG1700+518.** The scattering wind launches vertically from the accretion disk over a range of radii approximately coincident with the region emitting the bulk of the Balmer line emission. The emission disk and wind are rotating in the sense indicated by the symbols (dotted circle, motion out of the plane of the diagram; crossed circle, motion into the plane of the diagram). Radiation emitted by the accretion disk is scattered in the cylindrical inner section of the wind (shaded), which rises to ~$4r_{in}$, where $r_{in}$ is the inner radius of the broad emission-line region. The rotational and vertical components of the wind's velocity are both comparable with the local keplerian velocity. The broad absorption lines (BALs) are formed in the outer section of the wind, which is shielded from the high-energy radiation of the central AGN continuum by the scattering wind. The acceleration of this outer wind is dominated by line absorption of ultraviolet radiation from the central source, causing the flow to diverge along radial streamlines.